\documentclass{emulateapj}
\usepackage{epsf}
\begin{document}

\slugcomment{Accepted to ApJL January 10, 2012}
\shortauthors{A. L. King et al.}
\shorttitle{IGR J17091$-$3624}

\title{An Extreme X-ray Disk Wind in the Black Hole Candidate IGR J17091$-$3624}

\author{A.~L.~King\altaffilmark{1},
        J.~M.~Miller\altaffilmark{1},
        J.~Raymond\altaffilmark{2},
        A.~C.~Fabian\altaffilmark{3},       
        C.~S.~Reynolds\altaffilmark{4},
        T.~.R.~Kallman\altaffilmark{5},
        D.~Maitra\altaffilmark{1},   
        E.~M.~Cackett\altaffilmark{3,}\altaffilmark{6},
       M.~P.~Rupen\altaffilmark{7}}
       
\altaffiltext{1}{Department of Astronomy, University of Michigan, 500
Church Street, Ann Arbor, MI 48109-1042, USA, ashking@umich.edu}

\altaffiltext{2}{Smithsonian Astrophysical Observatory, 60 Garden Street, Cambridge, MA, 02138, USA}

\altaffiltext{3}{Institute of Astronomy, University of Cambridge,
  Madingley Road, Cambridge, CB3 OHA, UK}

\altaffiltext{4}{Department of Astronomy, University of Maryland,
  College Park, MD, 20742, USA}

\altaffiltext{5}{Laboratory for High Energy Astrophysics, NASA Goddard Space Flight Center, Code 662, Greenbelt, MD 20771, USA}

\altaffiltext{6}{Wayne State University, Department of Physics and Astronomy, Detroit, MI, 48201, USA}

\altaffiltext{7}{National Radio Astronomy Observatory, Socorro, NM 87801, USA}

\keywords{X-rays: binaries --- accretion, accretion disks --- black hole physics}

\label{firstpage}

\begin{abstract}
{\it Chandra} spectroscopy of transient stellar-mass black holes in
outburst has clearly revealed accretion disk winds in soft,
disk--dominated states, in apparent anti-correlation with relativistic
jets in low/hard states.  These disk winds are observed to be highly
ionized, dense, and to have typical velocities of $\sim$1000~km/s or
less projected along our line of sight.  Here, we present an analysis
of two {\it Chandra} High Energy Transmission Grating spectra of the
Galactic black hole candidate IGR J17091$-$3624 and contemporaneous
EVLA radio observations, obtained in 2011.  The second {\it Chandra}
observation reveals an absorption line at 6.91$\pm$0.01~keV;
associating this line with He-like Fe XXV requires a blue-shift of
$9300^{+500}_{-400}$~km/s (0.03$c$, or the escape velocity at 1000~R$_{Schw}$).  This projected outflow velocity is an
order of magnitude higher than has previously been observed in
stellar-mass black holes, and is broadly consistent with some of
the fastest winds detected in active galactic nuclei.  A potential
feature at 7.32 keV, if due to Fe XXVI, would
imply a velocity of $\sim 14600$~km/s (0.05$c$), but
this putative feature is marginal.  Photoionization
modeling suggests that the accretion disk wind in IGR J17091$-$3624
may originate within 43,300 Schwarzschild radii of the black
hole, and may be expelling more gas than accretes. The 
contemporaneous EVLA observations strongly indicate that jet activity
was indeed quenched at the time of our {\it Chandra} observations.  We
discuss the results in the context of disk winds, jets, and basic
accretion disk physics in accreting black hole systems.
\end{abstract}

\section{Introduction}
A detailed observational account of how black hole accretion disks
drive winds and jets remains elusive, but the combination of high
resolution X-ray spectroscopy, improved radio sensitivity, and
comparisons across the black hole mass scale hold great potential.
The broad range in X-ray luminosity covered by transient stellar-mass
black holes makes it possible to trace major changes in the accretion
flow as a function of the inferred mass accretion rate; this is
largely impossible in supermassive black holes.  Disk winds and jets,
for instance, appear to be state-dependent and mutually exclusive in
sources such as H 1743$-$322 \citep{Miller06, Blum10}, GRO J1655$-$40 \citep{Miller08,Luketic10,Kallman09}, and GRS 1915$+$105 \citep{Miller08, Neilsen09}.  This may offer insights into why many
Seyfert AGN, which are well known for their disk winds, are typically
radio--quiet \citep[though not necessarily devoid of jets; see][]{King11, Jones11,Giroletti09}.

The proximity of Galactic black hole binaries (BHB) ensures a high flux
level and spectra with excellent sensitivity in the Fe K band.  This
is of prime importance because He-like Fe XXV and H-like Fe XXVI lines
can endure in extremely hot, ionized gas \citep[see, e.g.][]{Bautista01}, and therefore trace the wind region
closest to where it is launched near the black hole.
Studies of some stellar-mass black hole disk winds find that the gas
is too ionized, too dense, and originates too close to the black hole
to be expelled by radiative pressure or by thermal pressure from
Compton heating of the disk, requiring magnetic pressure \citep{Miller06, Miller06b, Kubota07}.  Winds that may originate close
to the black hole and carry high mass fluxes are also observed in AGN
\citep[e.g.,][]{Kaspi02, Chartas02,King11b,Tombesi10}.

In this Letter, we present evidence of a particularly fast disk wind
in the black hole candidate IGR J17091$-$3624. The current outburst of IGR J17091$-$3624 was
first reported on 2011 January 28 \citep{Krimm11}.
Our observations caught IGR J17091$-$3624 in the high/soft state, but
it is important to note that the source has also showed low/hard state
episodes with flaring and apparent jet activity in radio bands  \citep{Rodriguez11}. X-ray flux variations
in IGR J17091$-$3264 bear similarities to the microquasar GRS 1915$+$105 \citep[e.g.,][]{Altamirano11}.

\section{Observation and Data Reduction}
\begin{figure}[t!]
\label{cont}
\includegraphics[scale=0.6,angle=-90]{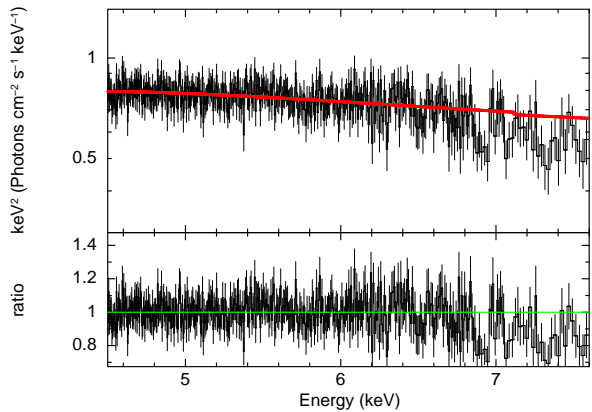}
\figcaption[t]{\footnotesize The second {\it Chandra}/HETG spectrum of
  IGR J17019$-$3624 is shown above, fit with a simple disk blackbody
  plus power-law continuum.  The continuum fit excluded the Fe K band
  to prevent being biased by line features.  The line at 6.91~keV is
  clearly apparent in the data/model ratio.  Associating this line
  with He-like Fe XXV implies an outflow velocity of $9300^{+500}_{-400}$km/s.  Weak evidence of a line at 7.32 keV, plausibly
  associated with Fe XXVI, would imply an even higher outflow
  velocity.  The data were binned for visual clarity.}
\end{figure}
IGR J17091$-$3624 was first observed with {\it Chandra} on 2011 August
1 (ObsID 12405), starting at 06:59:16 (UT), for a total of 30~ksec.
The High Energy Transmission Gratings (HETG) were used to disperse the
incident flux onto the Advanced CCD Imaging Spectrometer spectroscopic
array (ACIS-S).  To prevent photon pile-up, the ACIS-S array was
operated in continuous clocking or ``GRADED\_CC'' mode, which reduced
the nominal frame time from 3.2 seconds to 2.85 msec. The zeroth order flux is incident on the S3
chip, and frames from this chip can be lost from the telemetry stream
if a source is very bright.  We therefore used a gray window over the
zeroth order aimpoint; only one in 10 photons were telemetered within
this region.  For a longer discussion of this mode, please see, e.g.,
Miller et al.\ (2006) and Miller et al.\ (2008).  The source was
observed for a second time on 2011 October 6, starting
at 11:17:02 (UT), again for a total of 30~ksec.  The relatively low
flux observed during the first observation indicated that the ACIS-S
array could be operated in the standard ``timed event'' imaging mode
during this second observation.

Data reduction was accomplished using CIAO version 4.1 \citep{Fruscione06}.  Time-averaged
first-order High Energy Grating (HEG) and Medium Energy Grating (MEG) spectra were extracted from the Level-2 event
file.  Redistribution matrix files (rmfs) were generated using the
tool ``mkgrmf''; ancillary response files (arfs) were generated using
``mkgarf''.  The first-order HEG spectra and responses were combined
using the tool ``add\_grating\_orders''.  The spectra were then
grouped to require a minimum of 10 counts per bin.  All spectral
analyses were conducted using XSPEC version 12.6.0.  All errors quoted
in this paper are 1$\sigma$ errors.

Nearly simultaneous radio observations were made with the EVLA at each
{\it Chandra} pointing.  The first radio epoch included a two hour
integration at 8.4 GHz on 2011 August 2 (MJD 55776) at 1:01:04 (UT),
while the second was a two hour integration at both 8.4 and 4.8 GHz on
2011 October 6 (MJD 55841) at 22:10:16 (UT).  The flux and bandpass
calibrator was 3C 286.  The phase and gain calibrators were
J1720-3552 and J1717-3624, for the first and second observations,
respectively. The data are reduced using CASA version 3.2.1 \citep{McMullin07}.

\begin{figure}[!tp]
\label{xstar}
\includegraphics[scale=0.6,angle=-90]{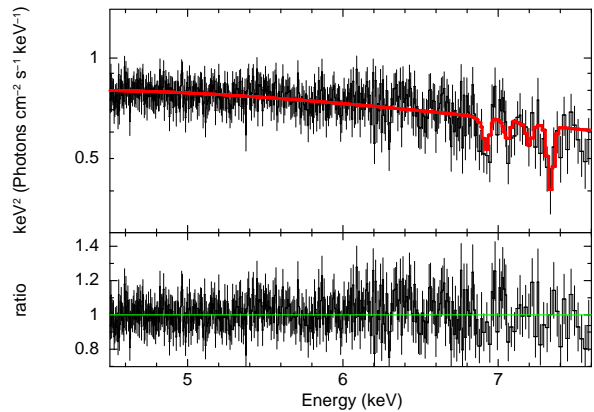}
\figcaption[t]{\footnotesize The second {\it Chandra}/HETG spectrum of
  IGR J17019$-$3624 is shown above, fit with a simple disk blackbody plus power-law continuum.  A self-consistent photoionization model, generated using XSTAR, was used to model the absorption in the Fe K band.  The data were binned for visual clarity.}
\end{figure}

\section{Analysis and Results}
A black hole mass has not yet been determined for J17091$-$3624; a value of $10~{\rm M}_{\odot}$ is assumed
throughout this work. Preliminary fits to the HETG spectra of IGR J17091$-$3624
suggested a relatively high column density, in keeping with
values predicted from radio surveys \citep[e.g. ${\rm N}_{\rm H} = 7.6
\times 10^{21}~ {\rm cm}^{-2}$; ][]{Dickey90}.
Due to this high column that predominantly affects lower energies, the MEG spectra have comparatively low sensitivity as compared to the HEG, and were therefore excluded.

The limitations of the HEG and our instrumental configuration enforce
an effective lower energy bound of 1.3~keV.  In the second
observation, the instrumental configuration served to enforce an upper
limit to the fitting range of 7.6~keV. This limit
was adopted for the first observation as well.

\subsection{The Spectral Continuum}
The HEG spectra were fit with a fiducial spectral
model including an effective H column density ({\it TBabs}), a disk
blackbody component, and a power-law component.

The first observation (MJD 55775) is well described by column
density of N$_H$=9.9$\pm 0.1 \times 10^{21}$ cm$^{-2}$, and a disk
blackbody temperature of 1.3$\pm0.1$ keV. The resulting fit gave a
$\chi^2/\nu$ = 2657/3156 = 0.84. This spectrum is dominated by the
disk black body component, typical of the high soft state of BHB.  A power-law continuum component is not
statistically required. An unabsorbed flux of F$_{2-10 keV}$ = 1.5$\pm 0.1\times10^{-9}$ ergs cm$^{-2}$ s$^{-1}$ was measured.

The second observation (MJD 55841) also had a consistent flux,
F$_{2-10 keV}$ = 1.9 $\pm 0.5 \times10^{-9}$ ergs cm$^{-2}$ s$^{-1}$.
Again, the column density was large, at N$_H$= 1.22$\pm 0.07 \times 10^{22}$ cm$^{-2}$.  A
power-law photon index of $\Gamma$= 1.7$^{+0.07}_{-0.09}$ and a disk
blackbody temperature of 2.3$\pm0.3$ keV were measured.  This disk
temperature is high but common in GRS 1915$+$105 \citep[see,
e.g.,][]{Vierdayanti10}.  The resulting $\chi^2/ \nu$ was 2754/3414=0.81.  

\subsection{The Line Spectra}
In the second HEG spectrum, absorption features are noted in the Fe K band 
(See Figure 1), and these were initially fit with simple Gaussians.  The two strongest lines are found at energies of
6.91$\pm0.01$ keV and 7.32$^{+0.02}_{-0.06}$. Via an F-test, \citep[see][for some cautions]{Protassov02}, these lines are significant at the
99.94\% and 99.67\% confidence levels respectively.  Dividing the flux
normalization of each line by its minus-side error suggests that
the feature at 6.91~keV is significant at the 4$\sigma$
level of confidence, while the 7.32~keV line is marginal at a 2$\sigma$ confidence level.

We also modeled the second observation continuum with a
Comptonization model ({\it compTT}) instead of the disk blackbody and
power-law. In general, this gave a reasonable fit
at $\chi^2 / \nu$ =2884/3414 = 0.84.  This model also showed residual
absorption features at high energy, which again we modeled with
Gaussian functions.  Relative to this continuum, the features at 6.91~keV
and 7.32~keV are detected at a higher level of significance
(6$\sigma$). 

It is reasonable to associate the line at $6.91\pm 0.01$~keV with
He-like Fe XXV, which has a rest energy of 6.70~keV \citep{Verner96}.
This translates into a blue-shift of 9300$^{+500}_{-400}$ km/s.  This
feature clearly signals an extreme disk wind in IGR J17091$-$3624. Typical velocities in X-ray Binaries are $<1000$~km/s \citep{Miller06,Miller06b}. If the feature at 7.32$^{+0.02}_{-0.06}$ keV is real and can be
associated with H-like Fe XXVI at 6.97~keV, it would correspond to a
blue-shift of 14600$^{+2500}_{-800}$.  For additional details, see
Table \ref{lines}. Although less likely, the 6.91~keV line could also be associated with a redshift from the H-like Fe XXVI line. The corresponding inflowing velocity would be 2600$\pm400$~km/s. If this is due to gravitational redshift, the corresponding radius would be 1.7$^{+0.3}_{-0.2}\times10^{8}$ cm (60$\pm10 ~R_{Schw}$). 

The absence of emission lines in the second spectrum of IGR
J17091$-$3624 is notable, but is only suggestive of an equatorial wind.
Given that disk winds have only been detected in sources viewed at
high inclination angles, and given the similarities between IGR
J17091$-$3624 and GRS 1915$+$105, it is likely that IGR J17091$-$3624
is also viewed at a high inclination. However, there is no evidence of eclipses in this source, so inclinations above 70$^\circ$ can be ruled out. 

Absorption lines like those detected in the second observation of IGR
J17091$-$3624 are absent in the first observation.  Fits to the Fe K
band using a local continuum model, and narrow Gaussian functions with
FWHM values corresponding to those measured in the second observation
give 1$\sigma$ confidence limits of 3~eV or less on lines in the
6.70--7.32~keV band.  This is significantly below the equivalent
widths measured in the second observation. (see Table 1) This may simply be due to
a variable absorption geometry in IGR J17091$-$3624; this has
previously been observed in H 1743$-$322 and GRS 1915$+$105 \citep{Miller06,Miller06b,Miller08,Neilsen09}.

\subsection{Photoionization Modeling}
To get a better physical picture of the absorption in the second
observation of IGR 17091$-$3624, we also fit the data with a grid of
self-consistent photoionization models created with XSTAR
\citep{Kallman01}. The ionizing luminosity for this model was derived 
from extrapolating the unabsorbed spectrum from the second observation to
 0.0136--30 keV, ensuring coverage above 8.8 keV, 
which is required to ionize Fe XXV. A distance of 8.5~kpc is first assumed to derive this luminosity (L$_{ion}$=3.5$\times10^{37}$ ergs/s), owing to the location of J17091$-$3624 within the Galactic bulge. However, \cite{Altamirano11b} also suggest the possibility that this source could be accreting at high Eddington fractions but further away, and a distance 
of 25~kpc was also adopted in a second XSTAR grid (L$_{ion}$=3.5$\times10^{38}$ ergs/s). 

The density of the absorbing material was chosen to be log($n$) = 12.0. This is a
reasonable assumption based on the modeling of similar X-ray binaries: GX 13+1,
$n = 10^{13}$ cm$^{-3}$\citep{Ueda04}, GRO J1655-40, $n = 10^{14}$
cm$^{-3}$\citep{Miller08}, H1743-322, $n=10^{12}$ cm$^{-3}$
\citep{Miller06}.  A turbulent velocity of $1000$~km/s was found to provide the best fit after various trials. A covering factor of 0.5 was chosen as the absence of emission lines suggests an equatorial
wind.  Finally, the Fe abundance was assumed to be twice the solar
value after initial fits; this characterizes the Fe K lines but does
not predict absorption lines, e.g. Si, that are not observed.

The initial, lower luminosity grid was fit to the data in XSPEC as a multiplicative model;
free parameters included the column density, ionization, and velocity
shifts of the absorbing gas (see Table \ref{lines} and Figure 2).  
For the disk blackbody and power-law continuum, an ionization parameter 
of $\log \xi$ = 3.3$^{+0.2}_{-0.1}$ is required, as well as a wind
column density of N = 4.7$^{+1.7}_{-1.9}\times 10^{21}$ cm$^{-2}$.  Velocity
shifts consistent with simple Gaussian models are found using the
XSTAR grid.

To fit the putative higher energy absorption, a second outflow
component is required.  An additional, lower luminosity XSTAR component is significant at the
$3\sigma$ level, relative to both continua.  The wind column density
was higher at N = 1.7$^{+1.2}_{-0.8} \times 10^{22}$ cm$^{-2}$, and the $\log \xi$ =
3.9$^{+0.5}_{-0.3}$. This system is moving even faster at 15400$\pm400$ km/s =
0.05c. (See Table \ref{lines} and Figure 2)

Repeating this analysis, but utilizing the higher luminosity XSTAR grid, we find that the two components are again required. In fact, the values of the column density, ionization and velocity shifts are nearly identical and well within $1\sigma$ of the previous model. 

To derive one estimate to the radius where these winds are
launched, we can estimate the radius at which the observed velocity equals the escape velocity. This constrains the radius to be at $r\simeq2.9 \times10^9$ cm (970 R$_{Schw}$). Using $\xi = L/(nr^2)$ and N$=nrf$, where f is the 1-dimensional filling factor, we can then derive the filling factor and density of the region. Assuming the ionizing luminosity is $3.5 \times 10^{37}$ erg/s, the resulting filling factor is $f \simeq 0.0008$, and the density is $n \simeq 2 \times10^{15}$ cm$^{-3}$. However, if the luminosity is higher (L$_{ion}$=$3.5 \times 10^{38}$ erg/s), the filling factor decreases to $f\simeq8\times10^{-5}$, and the density increases to $n\simeq 2\times10^{16}$ cm$^{-3}$.

These density estimates are quite high when compared to other X-ray binaries \citep[e.g.,][]{Ueda04,Miller06,Miller08}. However, we can invert the previous argument and instead derive the filling factor and radius from an assumed density, i.e. $n=10^{12}$cm$^{-3}$. We find a larger filling factor, ($f\simeq0.04$), and radius, ($r \simeq 1.3 \times 10^{11}$ cm, 43,300~$R_{Schw}$), if we require a luminosity of $3.5 \times 10^{37}$ erg/s. A larger luminosity, i.e. L$_{ion}$=$3.5 \times 10^{38}$ erg/s, reduces the filling factor,  ($f\simeq0.01$) but increases the radius ($r \simeq 3 \times 10^{11}$ cm, 100,000~$R_{Schw}$). At these radii the escape velocity is much lower than the observed velocity.

Finally, we can estimate the mass outflow rate ($\dot{m}_{wind}$) using a modified
spherical outflow, which can be approximated as $\dot{m}_{wind} \simeq
1.23 m_{p} L_{ion} f v \Omega / \xi$.  Here, we assume a covering factor $\Omega / 4\pi= 0.5$, and an outflowing velocity of $v = 9,600$~km/s. A luminosity of $L_{ion}= 3.5 \times 10^{38}$~erg/s and filling factor of $f=8\times10^{-5}$, gives a lower limit of $\dot{m}_{wind} \simeq 3.5 \times
10^{16}~ (10^{4}/\xi)$ g/s. However, a much larger outflow rate of  $\dot{m}_{wind} \simeq 1.7 \times
10^{18}~ (10^{4}/\xi)$ g/s is found, if we assume $L_{ion}= 3.5 \times 10^{37}$~erg/s and filling factor of $f=0.04$

For comparison, $L = \eta \dot{m}_{acc} c^{2}$, where $\eta$ is an
efficiency factor typically taken to be 10\%. For IGR J17091$-$3624,
$\dot{m}_{acc} = 3.8\times 10^{17}$ g/s. Using $\log \xi = 3.3$ from the
disk blackbody and power-law model, we find that the observed portion
of the outflow is likely to carry away 0.4--20 times the amount of accreted gas. Unless a geometrical consideration serves to bias our estimates, a high fraction of the available gas may not accrete onto the black hole. This trend is not only seen in BHB but in  Seyferts as well. \cite{Blustin05} note that more than half of their observed Seyferts show $\dot{M}_{out} / \dot{M}_{acc} > 0.3$ .

\subsection{Radio Non-Detections} 
The EVLA radio observations at 8.4 GHz were made nearly contemporaneously
with their X-ray counterparts.  Both radio observations were nearly two hours in duration.
Neither observation detected a source at the location of IGR
J17091$-$3624.  The RMS noise level for each observation was 0.02 mJy
and 0.07 mJy for the two epochs, respectively. The second observation had
extended coverage to 4.8 GHz that also had a non-detection.  The
RMS for this frequency was 0.13 mJy.  In contrast, IGR J17091$-$3624
was detected at the 1--2 mJy level during the low/hard state \citep{Rodriguez11}.  This supports prior findings that the
radio jet is absent during the periods when winds are seen in BHB
\citep{Miller06, Miller08, Neilsen10}.

\section{Discussion and Conclusions}
At ionizations above $10^{3}$, radiation pressure is inefficient, and
it is not able to drive these winds \citep[e.g.,][]{Proga00b}.  Thus, although the
UV components of disk winds in AGN are driven at least
partially by radiation pressure, the wind in IGR J17091$-$3624 likely
cannot be driven in this way.  A thermal wind can arise at radii
greater than $0.2~R_{C}$ \citep{Woods96}, where $R_{C} = (1.0\times
10^{10})\times (M_{BH}/M_{\odot})/T_{C8}$, where $T_{C8}$ is the
Compton temperature of the gas in units of $10^{8}$~K.  The spectrum
observed in the second observation gives $R_{C} \simeq 5\times
10^{12}$~cm. Therefore, if we assume our conservative estimate of the launching radius, it is possible for IGRJ17091$-$3624 to have a thermally driven wind. However, if the wind originates closer to the black hole, then it is likely that magnetic processes -- either pressure from magnetic viscosity within the disk \citep[e.g.,][]{Proga03} or
magneto-centrifugal acceleration \citep[e.g.,][]{Blandford82} -- must play a role in launching the wind observed in IGR J17091$-$3624.

Fast X-ray disk winds are not only seen in BHB like IGR J17091$-$3624,
but also in AGN and quasars \citep[e.g.,][]{King11b, Chartas02}.
The fastest UV winds observed in AGN are pushed to high
velocities by radiation pressure.  It remains to be seen whether a
common driving mechanism works across the black hole mass scale to
drive fast, highly ionized X-ray disk winds.  \cite{Chartas02} show
that in the quasar APM 08279+5255 there are broad absorption features,
which are likely highly relativistic Fe XXV and/or Fe XXVI lines.
In these regards, it bears some similarities to the most extreme winds
in BHB's.

Observations of BHB point to an anti-correlation
of wind and jet outflows from accretion disks \citep{Miller06,Miller08, Blum10, Neilsen10}.  Winds appear to only
be detected, or at least are considerably stronger, in soft,
disk--dominated states, and absent in hard states where compact,
steady jets are ubiquitous \citep{Fender06}.  In H 1743$-$322, in
particular, there is evidence that the absence of winds in hard states
is {\it not} an artifact of high ionization hindering the detection of
absorption lines, but instead represents a real change in the column
density (and thus the mass outflow rate) in any wind \citep{Blum10}.

It appears that our coordinated {\it Chandra} and EVLA observations of
IGR J17091$-$3624 support this anti-correlation.  The EVLA
observations place very tight limits on the radio flux when the disk
wind is detected, orders of magnitude below the level at which IGR
J17091$-$3624 was detected in radio during its low/hard
state only a few months prior \citep{Rodriguez11}.  

\cite{Neilsen09} suggested that the production of winds may be responsible for quenching jets in GRS 1915$+$105.  It
might then be the case that jets should be observed whenever winds are
absent.  In our first observation of IGR J17091$-$3624, however,
neither a wind nor a jet is detected, with tight limits.  Instead, the
apparent dichotomy between winds and jets may signal the magnetic
field topology in and above the disk is state-dependent.  This is
broadly consistent with multi-wavelength studies suggesting
synchrotron flares above the disk, but only in the hard state \citep[e.g. GX
339$-$4, XTE J1118$+$480,][]{DiMatteo99, Gandhi10}.  It is interesting to speculate that the magnetic field might be
primarily toroidal in the soft state, where a Shakura-Sunyaev disk is
dominant, but primarily poloidal in the hard state, when the mass
accretion rate is lower \citep[e.g.,][]{Beckwith08}. The type of outflow that is observed may
also depend greatly on how much mass is loaded onto magnetic field
lines; that could depend on variables including the mass
accretion rate through the disk. 

\vspace{0.2in}
We would like to thank the anonymous referee. We thank Michael Nowak for his instrumental help as well. ALK gratefully acknowledges support through the NASA
Earth and Space Sciences Fellowship.  JMM gratefully acknowledges
support through the {\it Chandra} Guest Observer program. The National Radio Astronomy Observatory is a facility of the National Science Foundation operated under cooperative agreement by Associated Universities, Inc.

\begin{deluxetable}{l l l l l}
\tablecolumns{5}
\tablewidth{0pc}
\tabletypesize{\scriptsize}
\tablecaption{Spectral Modeling Parameters of the 2$^{nd}$ HEG observation}
\tablehead{Parameter  & Model 1 & Model 2 & Model 3 & Model 4 \\ \hline
& diskbb + po & (diskbb + po) & comptt & (comptt) \\
& + Gauss + Gauss & $\times$ Xstar $\times$ Xstar  & + Gauss + Gauss  &  $\times$ Xstar $\times$ Xstar }
\startdata
N$_H$ (10$^{22}$ cm$^{-2}$) & 1.14$\pm0.06$ & 1.13$\pm0.06$ & 0.475 $^{+0.017}_{-0.018}$ & 0.558$^{+0.025}_{-0.028}$  \\
-\\
T$_{in}$ (keV) & 1.53 $\pm0.09$ & 1.51 $^{+0.11}_{-0.09}$& - & -\\ 
Norm & 13.1$^{+3.6}_{-2.5}$ & 13.8$^{+4.0}_{-1.7}$ & - &- \\
$\Gamma$ & 1.93 $^{+0.15}_{-0.16}$ & 1.91 $\pm0.17$& - & -\\
Norm &0.35 $^{+0.07}_{-0.08}$ & 0.34 $\pm0.08$ & - & -\\
-\\
T$_0$ (keV) & - &- & 0.58$\pm0.01$ & 0.59 $\pm0.01$  \\
kT (keV) & -& -& 10.5$^{+30}_{-1.7}$ & 9.8 $\pm0.02$ \\
$\tau_{plasma}$ & - & - & 2.24$\pm0.01$ &2.28 $\pm0.01$ \\
Norm & -&-& 0.0584$\pm0.0001$ & 0.062$^{+0.002}_{-0.05}$ \\
-\\
E$_{Fe XXV}$ (keV) & 6.91$\pm0.01$  & -& 6.91$^{+0.02}_{-0.01}$&-\\
FWHM (keV) & 0.091$^{+0.022}_{-0.049}$ & - & 0.13$^{+0.19}_{-0.04}$ &-\\ 
EW (keV)& 0.021$^{+0.005}_{-0.002}$ &  -&$0.040^{+0.007}_{-0.009}$&-\\
Norm ($10^{-4}$) & 3.5 $^{+0.8}_{-0.6} $& -&6.0$^{+1.1}_{-1.3}$ &-\\
v (km/s) & 9300$^{+500}_{-400}$ & -& 9300$^{+400}_{-800}$&-\\
-\\
E$_{Fe XXVI}$ (keV) &7.32$^{+0.02}_{-0.06}$& -& 7.30$\pm$0.02 &-\\
FWHM (keV) & 0.081$^{+0.079}_{-0.027}$ & - & 0.25 $^{+0.13}_{-0.01}$&-\\ 
EW (keV) & 0.032 $^{+0.018}_{-0.004}$&  -&$0.089^{+0.013}_{-0.014}$&-\\
Norm ($10^{-4}$) & 3.4$^{+1.9}_{-0.4}$ & -&11.8$^{+1.7}_{-1.6}$ &-\\
v (km/s) &14600$^{+2500}_{-800}$& -& 13800$\pm800$&-\\
-\\
N (10$^{22}$ cm$^{-2}$) & - & 0.47$^{+0.17}_{-0.19}$ & - & 0.45$^{+0.33}_{-0.17}$ \\
$\log \xi$(ergs cm s$^{-1}$) & - &  3.3 $^{+0.2}_{-0.1}$ &-&3.4 $^{+0.2}_{-0.1}$ \\
v (km/s) & - & 9600$^{+400}_{-500}$ & - & 9600$\pm300$ \\
-\\ 
N (10$^{22}$ cm$^{-2}$) & - & 1.66 $^{+1.18}_{-0.83}$ & - & 1.97$^{+1.26}_{-0.51}$ \\
$\log \xi$(ergs cm s$^{-1}$) & - &  3.9$^{+0.5}_{-0.3}$ &-&3.7 $^{+0.3}_{-0.1}$ \\
v (km/s) & - & 15400$\pm400$& - & 15400 $^{+400}_{-300}$\\
-\\
$\chi^2/\nu$ & 2725/3408 = 0.80 & 2731/3408 = 0.80 & 2793/3408 = 0.82 & 2761/3408 = 0.81 \\
 \enddata  
\label{lines}
\tablecomments{This Table lists the line detections using Gaussian functions as well as more self-consistent, photoionization components created with XSTAR, assuming two different continuum models. {\it TBabs} is applied to all the models and the errors are 1$\sigma$ confidence level.}
\end{deluxetable}

\clearpage



\end{document}